\documentstyle[epsfig,amstex]{aipproc}

\begin{document}
\title{Recent Photoproduction Results From ZEUS}
\author{R. Saunders\\ {\small for the ZEUS collaboration}}
\address{University College London, Gower St., London, WC1E 6BT, England}
\maketitle
\vspace{-7cm}
\begin{flushright}
UCL/HEP 97-01
\end{flushright}
\vspace{6cm}

\begin{abstract}
Recent results for inclusive jet cross sections, dijet cross sections and 
dijet angular distributions are compared with NLO perturbative QCD calculations.
The observation of isolated high $P_T$ photons 
(prompt photons) is also reported.  
\end{abstract}
\section*{introduction}
Collisions of 27.5 GeV positrons with 820 GeV protons at {\bf HERA} are 
studied using the {\bf ZEUS} detector.  The processes of interest here are
those where an almost real photon (virtually $P^2\approx0$) is radiated from the 
positron which then interacts with a parton in the proton to produce two 
high transverse energy jets.  The hard scale provided by the large transverse 
momentum allows us to describe these processes using perturbative QCD (pQCD), 
and therefore they can be used to test it.  The photon can either interact 
directly with the parton from the proton (direct processes), or it can first 
fluctuate into an hadronic state which then acts as source of partons for the 
hard scatter (resolved processes). In the resolved case, in addition to the two
jets, there is also a photon remnant.  In 
direct process since all the photon  takes part in the interaction 
$x_\gamma$, the fraction of the photon momentum entering into the hard scatter, is 1,
and conversely for resolved processes $x_\gamma<$1.  In order to use this to
identify and separate direct and resolved event samples it is necessary to
define an experimentally measurable $x_\gamma$ that is valid to all orders.
This is done in terms of the transverse energy $E_t^{jet}$ and
pseudorapidity $\eta^{jet}=-ln(tan\frac{\theta^{jet}}{2}))$ of the two
highest transverse energy jets in an event~\cite{ZEUS-hpp1}: 
\begin{equation}
x_\gamma^{OBS}=\frac{1}{2yE_e}\left(E_t^{jet1}e^{-\eta^{jet1}}+ E_t^{jet2}e^{-\eta^{jet2}}\right)
\end{equation} 
Higher order and hadronisation effects cause the average
$x_\gamma^{OBS}$ of direct events to be slightly less than 1, and the distribution peaks at $\sim$0.85. A 
cut at $x_\gamma^{OBS}=0.75$ is used to separate the two classes of event. 
Direct processes are sensitive to high $x_\gamma$ quarks in the photon, and the
parton distribution of the proton, whereas resolved process are sensitive
to lower $x_\gamma$ parton distributions in the photon.
The results presented here are based on 2.7pb$^{-1}$ of data taken in the 1994
HERA running period, except the prompt photon analysis which uses data coresponding to 5.7pb$^{-1}$
from 1995.  All cross sections presented here are corrected back to final state
particle level using Monte Carlo.
\section*{Inclusive jet cross sections}
\begin{figure}
\centering
\epsfig{file=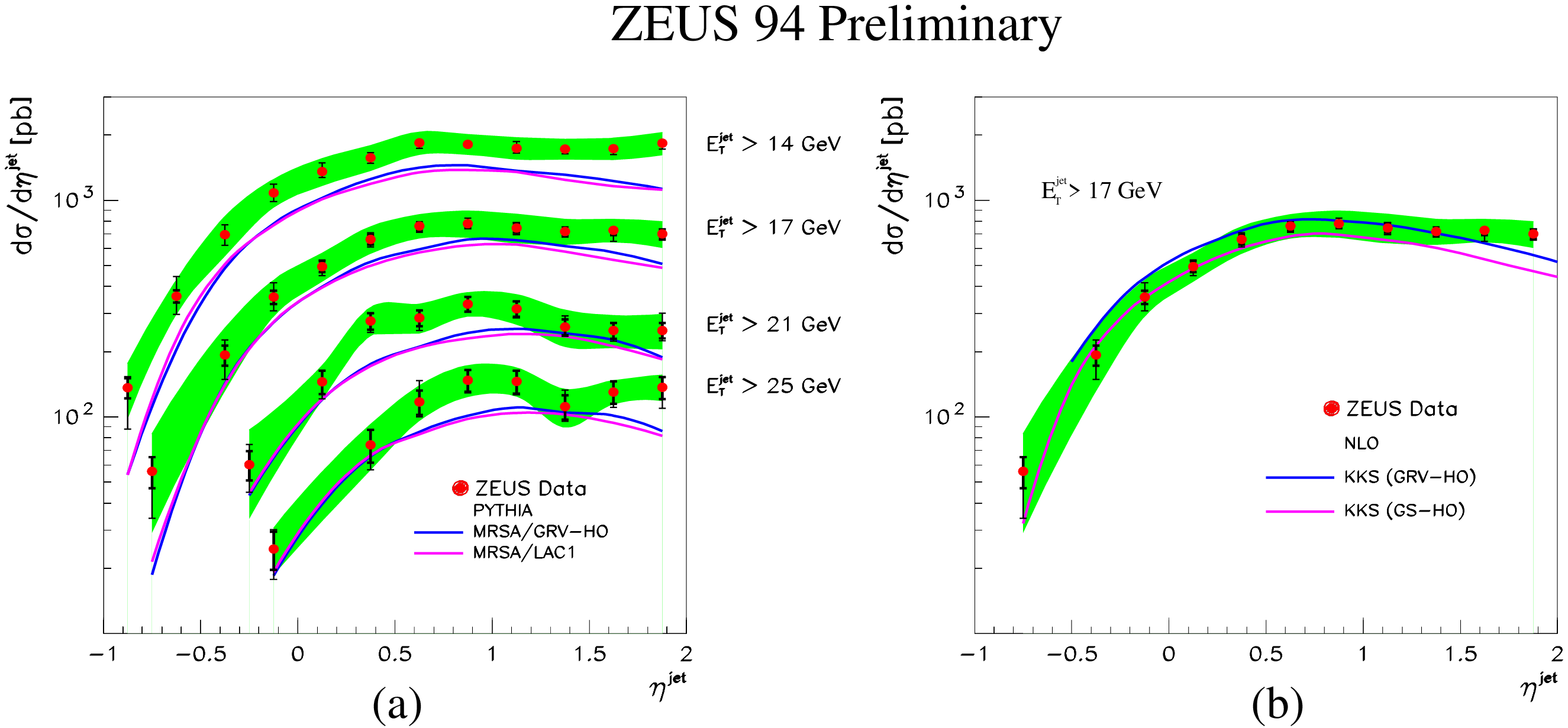, height=14cm, angle=90}
\vspace{-0.5mm}
\caption{\label{f-inc-xsecs} Inclusive jet cross sections. Thick error bars: statistical, thin error bars:systematic, shaded band: uncertainty due to the calorimeter energy scale.}
\end{figure}
\vspace{-5mm}
The cross section of the hard scatter is factorised into three parts:
the parton-parton scattering cross section $\hat\sigma(ab\rightarrow j_1j_2)$,
the parton densities in the proton $f_{a/p}(x_p,Q^2)$, and those of the 
photon $f_{a/\gamma}(x_\gamma,Q^2)$.  The jet cross section at leading order
is written as
\begin{equation}
\begin{split}
\frac{d\sigma(\gamma p \rightarrow j_1j_2X)}{dp^2_td\eta_1d\eta_2}& = \sum_a\sum_bx_px_\gamma f_{a/p}(x_p,Q^2) \\
&\times f_{b/\gamma}(x_\gamma,Q^2)\frac{d\hat\sigma(ab\rightarrow j_1j_2X)}{d\hat{t}}
\end{split}
\end{equation}
where $\eta_1$ and $\eta_2$ denote the  pseudorapidities of the two outgoing 
partons.  So by measuring jet cross sections one can test pQCD and gain 
insight into the parton densities in the proton and photon.  At 
HERA we are looking at e-p collisions and there is a spectrum of incoming 
photon energies, usually described using the Weizs\"acker-Williams 
approximation~\cite{ww}. Here the inclusive jet cross section 
$\frac{d\sigma}{d\eta^{jet}}$ is presented for
cone jets with a radius R=1 in $\eta$ - $\phi$ space, with pseudorapidities
$-1<\eta^{jet}<2$, in events with centre of mass energies 
134$< W=\sqrt{yE_eE_p}<277$ GeV.  The measurement is made for cuts on the 
minimum $E_t^{jet}$ of 14, 17, 21, and 25 GeV.  Figure~\ref{f-inc-xsecs}(a) shows 
these cross sections compared to expectations from the PYTHIA model.  PYTHIA
simulates the matrix element in LO with the inclusion of initial and final
state parton showers (PS).  The shape of the cross sections is well described
by the model, however the PYTHIA result is $\sim 30\%$ lower than the data in 
all cases.
This is true for both GRV-HO and LAC1 parameterisations of the photon structure function.
Figure~\ref{f-inc-xsecs}(b)  shows the curve for $E_t^{jet}>$17 GeV compared
to next to leading order (NLO) QCD calculations by Kramer, Klasen, and Salesch~\cite{kks},
using GRV-HO and GS parton distributions for the photon.
The calculated and measured cross sections agree well in magnitude up to $\eta^{jet}\approx$ 1.3 for both 
photon parton distribution functions.  At larger rapidities the calculations tend to fall below the data.

\section*{DIjet Cross sections}

The dijet cross sections as a function of 
$\bar{\eta}=\frac{1}{2}(\eta^{jet1}+\eta^{jet2})$ have been measured 
in order to study the photon and proton structure.
The requirement $|\eta^{jet1}-\eta^{jet2}|<$0.5 is made to increase
correlation to the parton momenta and ensure a good resolution for the
measurement.
The measurement was made with a clustering algorithm KTCLUS~\cite{ktclus},
and  two types of cone algorithm EUCELL and PUCELL(both with a cone radius 
of 1).  These algorithms differ in their 
seed finding and cone merging. Figure~\ref{f-dijet-2}(a) shows the measured
cross sections for all three jet finders for minimum $E_t^{jet}$ cuts
of 6, 8, 11, and 15 GeV.  The upper four plots are for $x_\gamma^{OBS}>0.75$,
and the lower four for $0.3<x_\gamma^{OBS}<0.75$.  To model the differences in
the jet algorithms in NLO calculations an additional parameter, $R_{sep}$, is 
introduced~\cite{rsep,bfkk}.  
Curves for
$R_{sep}=$1 and $R_{sep}=$2 are also shown in figure~\ref{f-dijet-2}(a).
$R_{sep}=1$ emulates KTCLUS and $R_{sep}=$1R ($\approx 1.5 - 2$R) PUCELL 
(EUCELL).
At lower $E_t^{jet}$ the differences between the two calculations and the 
different jet algorithms are similar. As the $E_t^{jet}$ rises the
differences between the measurements gradually disappear, however the
differences between the calculations remain.
For the high $x_\gamma^{OBS}$ cross sections there is good
agreement between the measurements and the respective calculation.  
The same is true
for the low $x_\gamma^{OBS}$ results with $E_t^{jet}>11$ GeV.
However, when lower energy
jets are included the data lie significantly above the calculations.  

\section*{Dijet angular distributions}

\begin{figure}
\centering
\epsfig{file=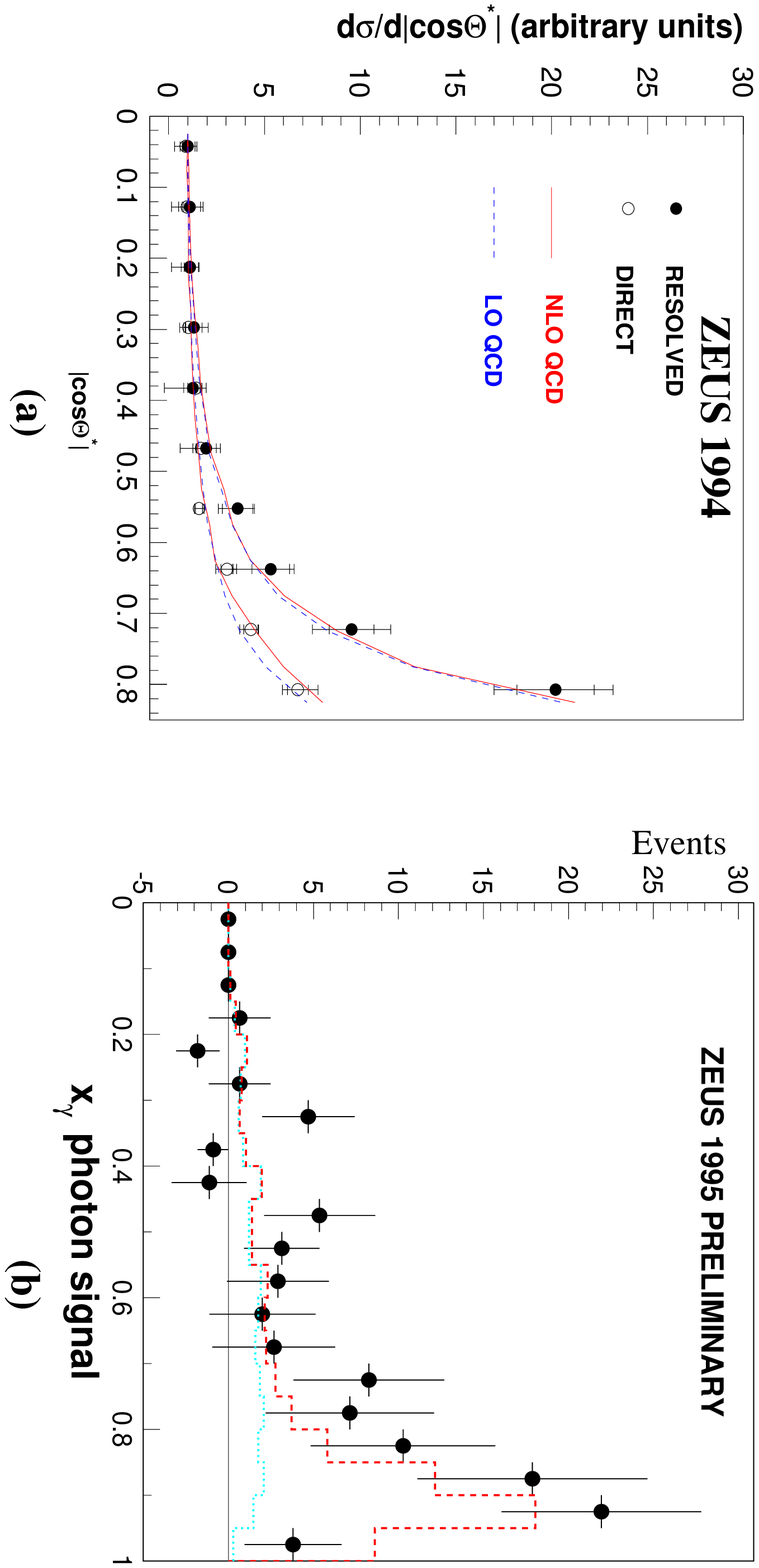,width=7cm, angle=90}
\vspace{-1mm}
\caption{\label{f-dijet-1} (a) Dijet angular distributions Compared to LO and NLO calculations. (b) Observed $x_\gamma$ distribution for prompt photon events compared to direct+resolved (Dashed line), and resolved only (Dotted line) PYTHIA samples.}
\vspace{-3mm}
\end{figure}

In it's CMS the dijet system is characterised by a scattering angle $\theta^*$,
the angle between the jets and the beam axis.  The dominant direct photon process is photon-gluon fusion $(\gamma g \rightarrow q \bar{q})$ where the exchanged particle is a quark. \nolinebreak This 
\begin{center}
\epsfig{file=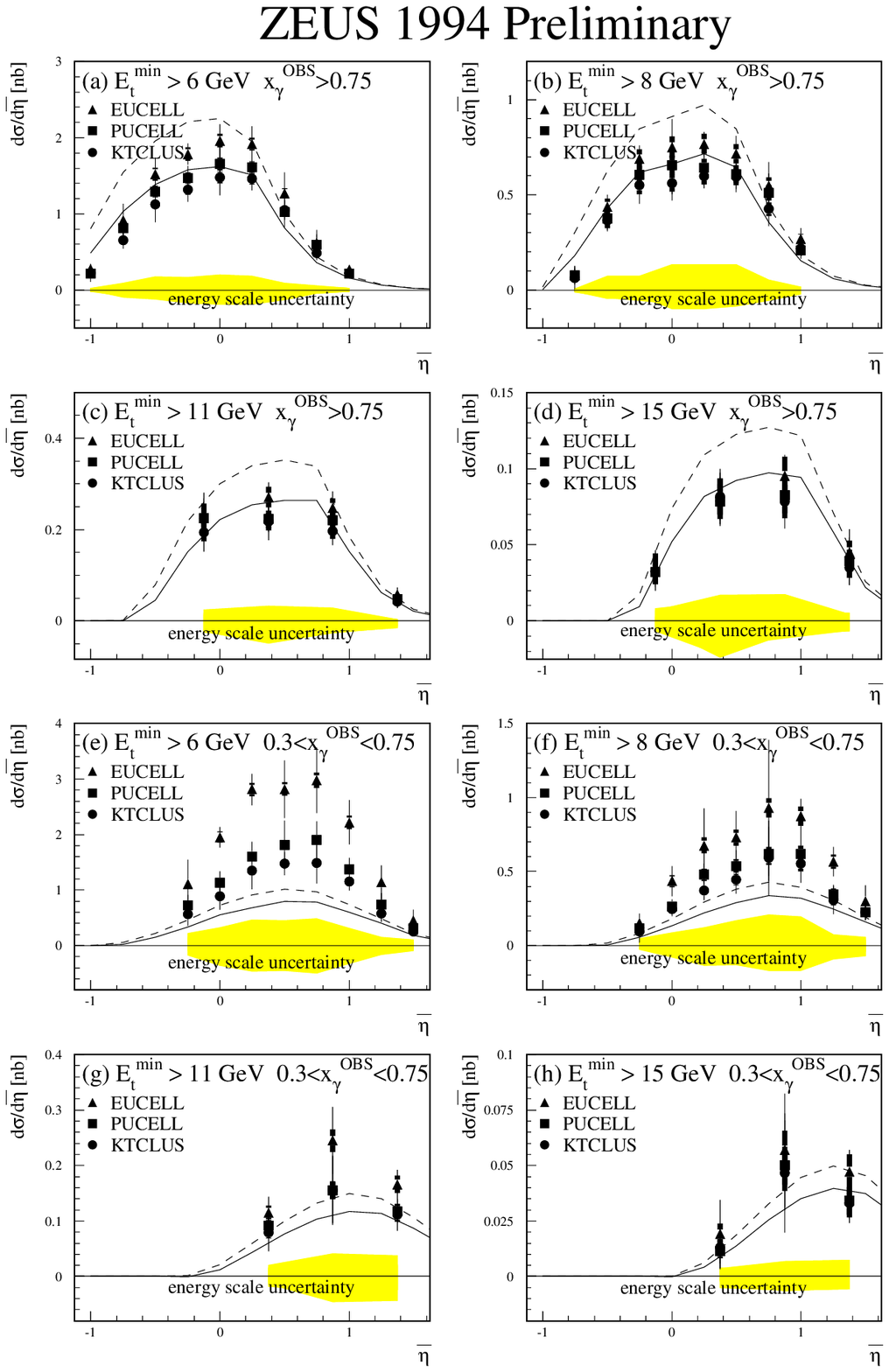,height=11cm}
\vspace{-3.8mm}
\begin{figure}[h]
\caption{\label{f-dijet-2} Dijet cross sections. Points: data using different jet algorithms, thin (thick) error bars:systematic (statistical), shaded band: calorimeter energy scale uncertainty.  The curves
are NLO calculations with $R_{sep}=1$ (solid) and $R_{sep}=2$ (dashed).}
\end{figure}
\end{center}
\vspace{-0.5cm}
fermion exchange leads to a $(1-|cos\theta^*|)^{-1}$ angular dependence
in the 
cross section.  In contrast resolved processes are dominated by gluon exchange,
which results in an angular dependence of $(1-| cos\theta^*|)^{-2}$.  Therefore low and
high $x_\gamma^{OBS}$ samples should have different angular distributions.  
Figure~\ref{f-dijet-1}(a) shows the measured $\frac{d\sigma}{d|cos\theta^*|}$ 
for direct and resolved event samples normalised to unity at
$|cos\theta^*|=0$.  
Also shown are curves for LO and NLO calculations~\cite{owens-et-al}.  
As expected, the resolved sample
 shows a much steeper rise towards  high $|cos\theta^*|$ than the
direct sample.  The data and calculations are in excellent agreement.  This confirms that
direct (resolved) photoproduction is dominated by the exchange of spin-$\frac{1}{2}$ quarks (spin-1 gluons). 

\section*{prompt photon} 

The dominant prompt photon processes are $qg\rightarrow\gamma q$ 
and $q\bar{q}\rightarrow\gamma g$ for resolved, and 
$\gamma q\rightarrow q \gamma$ for direct 
photoproduction.  
Therefore prompt 
photon measurements are particularly sensitive to the quark distributions in the 
proton and photon.  
Direct processes are expected to dominate overall.  The advantage of the photon in the final state
over jets is that it is not subject to hadronisation effects.  Events are selected if they contain at least
one jet with $E_t>$4 GeV and  one prompt photon.  The photons are identified as an electromagnetic 
cluster in the calorimeter with 5$<E_T<$10 GeV and no track within a radius of 0.3 in $\eta$ - $\phi$ space. 
Figure~\ref{f-dijet-1}(b) shows as points the measured $x_\gamma$ distribution for
these events.  Also shown are histograms for full, and resolved only PYTHIA Monte Carlo samples.  
This plot confirms that the direct processes dominate. 


\begin{references}
\bibitem{ZEUS-hpp1}ZEUS Collaboration, {\it Phys. Lett.}\ {\bf B 348}, 665-680 (1995).
\bibitem{kks}M. Klasen, G. Kramer and S. G. Salesch, {\it Z. Phys.}\ {\bf C 68}, 113 (1995).
\bibitem{ktclus} S.~Catani, Yu.L.~Dokshitzer, M.H.~Seymour and B.R.~Webber, \\
{\it Nucl.\ Phys.}\ {\bf B 406} 187 (1993).
\bibitem{rsep} S.D.~Ellis, Z.~Kunszt, D.E.~Soper, {\it Phys. Rev. Lett.}\ {\bf 69} 3615 (1992).
\bibitem{bfkk} J. M. Butterworth, L. Feld, M. Klasen and G. Kramer,
hep-ph 9608481;\\
{\it Proceedings of the workshop Future Physics at HERA} 554 (1996).
\bibitem{ww} C.F.v. Weizs\"acker, {\it Z. Phys.}\ {bf 88}, 612 (1934)\\
 E. J. Williams, {\it Phys. Rev.}, {\bf 45}, 729 (1934).
\bibitem{owens-et-al}H. Baer, J. Ohnemus and J.F. Owens {\it Phys. Rev.}\ {\bf D 40}, 2844 (1989); \\
J.F. Owens, private communication.
\bibitem{kk}M. Klasen, G. Kramer DESY-96-246, hep-ph/9611450.
\end{references}
\end{document}